# Blockchain in Global Supply Chains and Cross Border Trade: A Critical Synthesis of the State-of-the-Art, Challenges and Opportunities

Yanling Chang[1,2]*, Eleftherios Iakovou[1,3] and Weidong Shi[4]

*[1]Department of Engineering Technology & Industrial Distribution, Texas A&M University, College Station, TX, USA; [2]Department of Industrial Systems and Engineering, Texas A&M University, College Station, TX, USA; [3]Manufacturing and Logistics Innovation Initiatives, Texas A&M Engineering Experiment Station (TEES), College Station, TX, USA; [4]Department of Computer Science, University of Houston, TX, USA*

*Corresponding author. Email: yanling.chang@tamu.edu

Blockchain in supply chain management is expected to boom over the next five years. It is estimated that the global blockchain supply chain market would grow at a compound annual growth rate of 87% and increase from $45 million in 2018 to $3,314.6 million by 2023. Blockchain will improve business for all global supply chain stakeholders by providing enhanced traceability, facilitating digitisation, and securing chain-of-custody. This paper provides a synthesis of the existing challenges in global supply chain and trade operations, as well as the relevant capabilities and potential of blockchain. We further present leading pilot initiatives on applying blockchains to supply chains and the logistics industry to fulfil a range of needs. Finally, we discuss the implications of blockchain on customs and governmental agencies, summarize challenges in enabling the wide scale deployment of blockchain in global supply chain management, and identify future research directions.

## 1. Introduction

Successful global supply chain management (SCM) hinges upon the comprehensive and harmonized management of four flows, namely, the flows of products, processes, information, and cash. This concerted effort affects profoundly the competitiveness of a business in terms of product cost, working capital requirements, speed-to-market, service and eventually profitability. Technological developments over the past few decades have make it easier and faster to move cargo from one location to another across the globe. However, despite these advances, today's globalized supply chains (SCs) face challenges when it comes to tracing events and investigating incidents, ensuring the integrity of cargo, resolving disputes, digitalization, compliance, and enabling trust among the involved parties across complex SCs (see Figure 1). A recent study revealed that out of 408 organizations and corporations from 64 countries, 69% of them lack full visibility into their SCs, 65% of them experienced at least one SC disruption, and 41% of them still rely heavily on Excel spreadsheets for tracking SC disruptions (Microsoft 2018).

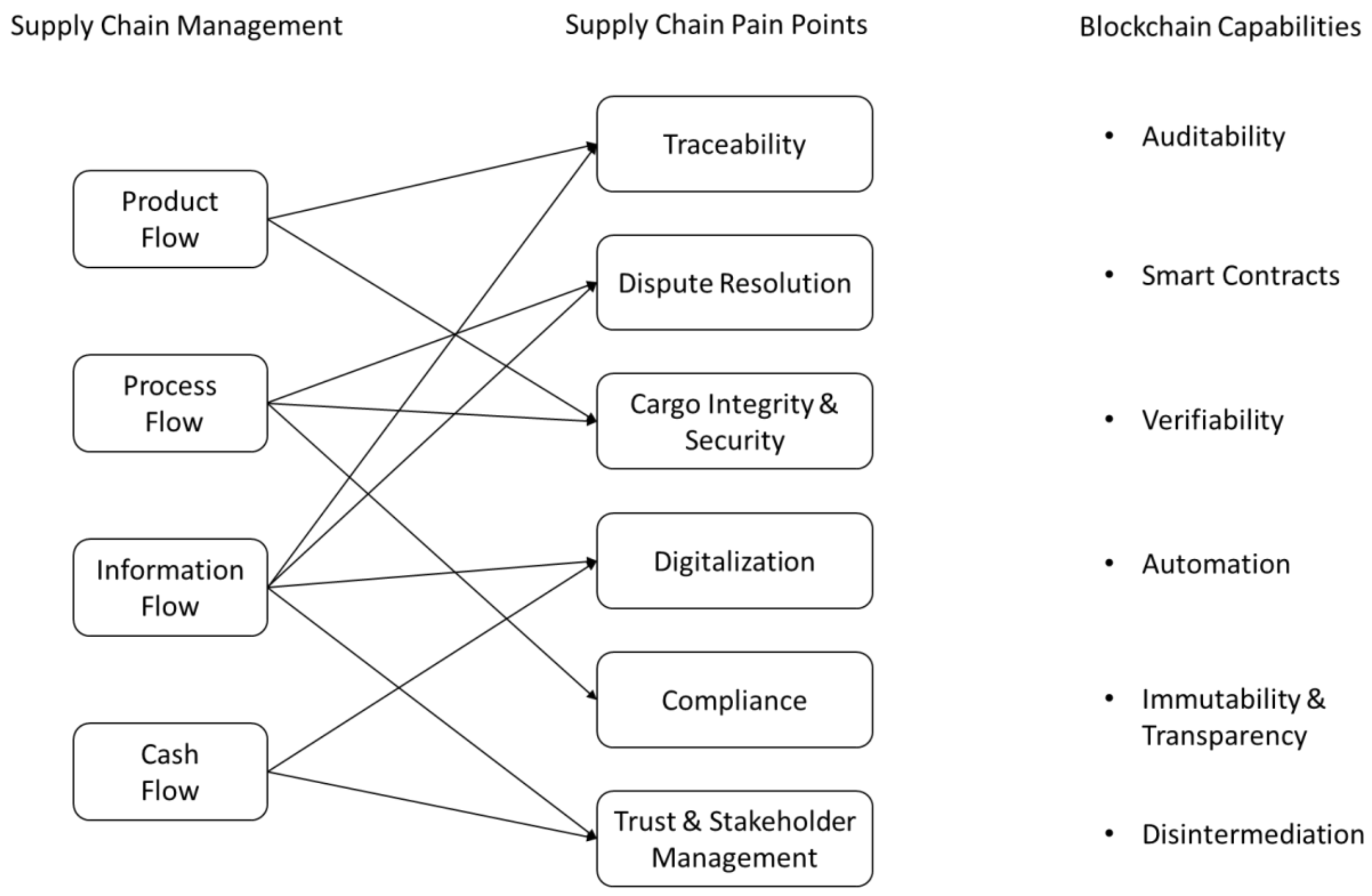

Figure 1: Supply chain pain points and capabilities of blockchain

In this paper, we first outline today's landscape and then provide a critical synthesis of the emerging challenges and pain points in global SCs and cross-border flows in Section 2. We then introduce the concepts of blockchain technology and smart contracts, and discuss their implications on global trade and SCM in Section 3. In Section 4, we provide a taxonomy of the leading pilot efforts as these are mapped according to the critical issues identified in global SCs. The implications of blockchain for customs and governmental agencies are discussed in Section 5. In Section 6, we outline the challenges for the wide adoption of blockchain in global trade and SCM, while we wrap-up in Section 7 with future research directions involving the participation of academia, industry and government.

## 2. Challenges for Today's Global Supply Chain Management

### *2.1 Traceability*

SC traceability refers to 'the ability to identify and trace the history, distribution, location and application of products, parts and materials, to ensure the reliability of sustainability claims, in the areas of human rights, labour (including health and safety), the environment and anti-corruption' (United Nations Global Compact 2016). Traditionally, traceability has focused mainly on upstream supply networks, tracking the source and origin of the raw materials and components. Nowadays, its scope has expanded to downstream capabilities, tracing goods along the multi-layer distribution networks through the end consumers (Supply Chain Digest 2017).

Traceability allows business stakeholders, authorities, governmental agencies and ultimately consumers, to manage and respond to risks in a responsive and documented way. Unfortunately, due to the endemic lack of traceability and transparency, customers and buyers often have no reliable and efficient way to verify and validate the sources and details about the products and services they purchase. Indicatively only, in 2015, 55 people were infected with an E.coli outbreak at Chipotle Mexican Grill outlets. Chipotle and the Centers for Disease Control and Prevention (CDC) were still unable to definitively identify the source of contamination after months of investigation (CDC 2016).

Provenance is an essential characteristic of what customers buy for many products. Fair-trade, non-GMO, and organic certifications are highly sought-after and can be easily faked. Indicatively, halal, kosher and organic foods are indistinguishable from their corresponding conventional products. Counterfeit products, especially in food and pharmaceutical chains, could be detrimental to both brand integrity and consumers. Consumers, governments, and companies are now increasingly demanding more transparency from brands, manufacturers, and producers throughout the SC. Recently, 70,000 consumers signed a petition that urged large companies and brands, including Walmart and Forever 21, to boost SC transparency (Scarano 2018).

Supply Chain Digest (2017) has identified the following drivers for expanded traceability capability:

- Regulatory requirements, legal frameworks and laws especially within life science and food industries.
- Ethical compliance and environmental sustainable practices of corporations.
- Security/safety for products vulnerable for theft or counterfeiting.
- History of physical flows and movement conditions of products and assets for complete, reliable and prompt recall operations.

### *2.2 Dispute Resolution*

The sustained flow of commerce along a complex global SC network will inevitably attract disputes. Many of these disputes are the result of poor contract management at the outset of supplier-buyer relationships. SC contracts are often drafted by the purchasing or contracting personnel in isolation, without much input and guidance from the operating units that perform the contract, primarily, engineering and finance. As a result, disputes can arise when participants begin contract performance without verifying with the SC contract's terms. Often, when an engaged participant fails to deliver required products on-time and in-full or the products have been compromised en route, SC stakeholders have to identify the problem quickly and usually settle the dispute by fines or compensation. For example, as Walmart is pushing for on-time, in full deliveries, it began fining suppliers who cannot deliver at least 85% of their shipments on time, in April of 2018 (Supply Chain Dive 2018).

SC disputes typically are difficult and expensive to pursue and manage, even if there is only a relatively small amount of money in dispute. Tracking back auditing in order to identify cause is both error-prone and costly. These disputes may undermine relationships that have been in place for years and could involve multiple business functions including engineering, contracting, quality assurance, finance, and executives from various organizations. Moreover, when a (cross-border) dispute arises, each party may be reluctant to enter the other’s courts to settle it. Involved stakeholders may not want to confront with protective judges; or they are unfamiliar with (and therefore skeptical of) local laws. How to quickly and efficiently resolve disputes when conditions listed on the contract are not met, is a challenging task.

### *2.3 Cargo integrity and security*

As cargo moves from upstream to downstream along a SC, two types of agents are involved: the owners, who have complete ownership rights of the cargo, and the carriers, who have just been delegated ownership rights. Thus, the product flow can also be viewed as a sequence of ownership transfer along the SC in which the current owner (seller) of the cargo transfers it to a subsequent owner (the buyer) via carriers.

Bills of lading, policies of insurance, and invoices are crucial documents used in international trade to ensure that buyers receive payment and sellers receive genuine and uncompromised cargos. However, these documents are not foolproof, as criminals or adversaries can steal everything from package deliveries to entire shiploads with fraudulent documents. Fraudsters can create a fake set of bills of lading and other cargo documents that look sufficiently genuine to take delivery of the cargo in advance of the actual recipient. Recently, the logistics and maritime commerce industries have experienced a sharp increase in the number and variety of fraud cases, including: fraudulent misrepresentations on cargo documents, sales of cargoes that do not exist, fake letters of indemnity, cheating over quantity and quality, pilferage, and cargo theft. CargoNet recorded 836 cargo thefts in the U.S. in 2016 with an average value of $207,000 with the total value of the stolen cargo amounting to $114 million. An estimated $30 to $50 billion worth of cargo is stolen worldwide per year (Transport Topics 2017).

Furthermore, the maritime industry is also exposed to a wide range of cyber risks, including cyber extortion, fraud and theft, as well as acts of cyber terrorism and piracy. In late 2013, the Belgian port of Antwerp reported that crime syndicates have used sustained cyber-attacks for drug trafficking while these attacks had been ongoing for more than two years undetected. Drug traffickers hid cocaine and heroin among legitimate cargos and recruited highly intelligent hackers to breach IT systems that controlled the movement and location of containers. The penetration allowed the traffickers to remotely access to the terminal systems, and thereby they were able to release containers to their own truckers. Furthermore, the access to port systems was used to delete information and created many ‘ghost containers’ (Seatrade 2013).

CyberKeel (Maritime cyber risk, www.cyberkeel.com) has mapped the flow of usual information exchanges from the point of booking a container until delivery at the endpoint – shipping lines, logistics companies, ports, terminals, customs authorities and IT data portals. The mapping showed more than 50 possible locations and points to be vulnerable to cyber-attacks. Often, penetration at one or two such points would be sufficient to allow or facilitate an unauthorized movement of goods and cargo.

As the global trade is becoming more reliant on IT and electronic trading platforms and documents, the risk of cyber-attacks and insider threats is increased. Understanding the interrelationship between cyber and physical transportation security is a new requirement and poses significant challenges to global trade and SCM.

### *2.4 Supply Chain Digitalization*

Ninety percent of goods in global trade are carried by the maritime industry each year (International Chamber of Shipping 2017). The vast number of containers traveling among the world's ports cannot complete their journey alone. They are associated with tons of paperwork and documents including bills of lading, packing lists, letters of credit, insurance policies, orders, invoices, sanitary certificates, certificates of origin, etc. Maersk, the global leader in transport and logistics, found in 2014 that a single shipment of refrigerated goods from Mombasa to Europe can generate 200 separate communications and interactions between nearly 30 organizations (IBM 2017). The stack of documents created by these 200 communications measured about 25 centimetres in height (Allison 2016). Ships or aircrafts are often delayed in ports because the paperwork has not caught up with the products they carry.

The cost of these administrative practices is enormous and the existing SC networks suffer from issues of data duplications, inconsistency and redundancies, etc. Processing trade documents can cost as much as a fifth of the cost of shifting goods. Removing administrative blockages and outdated practices in SCs could do more to boost international trade than eliminating tariffs. Reducing SC barriers to trade could increase GDP up to six times more than removing tariffs (World Economic Forum 2013). The UN reckons that full digitization of all the Asia-Pacific region's trade-related paperwork could reduce cost by up to 31% and boost exports by as much as $257 billion per year (The *Economist,* March 22nd, 2018; April 26th, 2018).

While the global maritime SC is still a legacy business, SC digitization will bring down these barriers and create a completely integrated ecosystem for improved efficiencies and transparency (Schrauf 2017). A recent McKinsey study (Bughin 2017) estimated that companies could raise annual growth of earnings before interest and taxes by 3.2 percent and improve the annual revenue growth by 2.3 percent by aggressively digitizing their SCs. However, the same study also showed that the current digitalization level of SCs is only 44 percent (see Figure 2), and merely 2 percent of the surveyed executives thought they should focus on digitalizing their SCs.

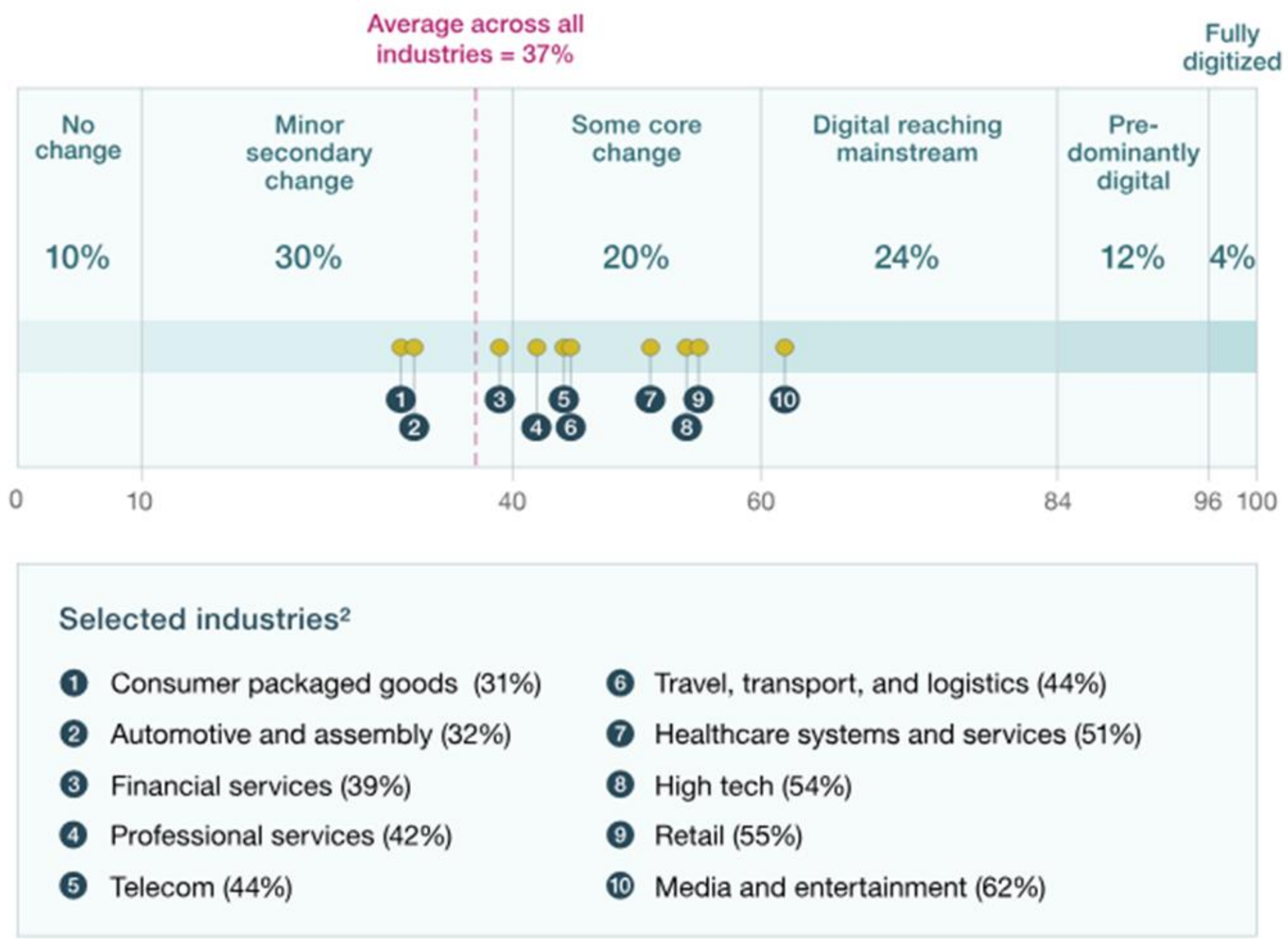

Figure 2: Perception of digital penetration by industry (Bughin 2017)

### *2.5 Compliance*

According to Deloitte (2015), SC compliance refers to 'organizational adherence to established guidelines and requirements that relate to each risk domain along the SC continuum, as well as to an organization's ability to meet or exceed the expectations of its stakeholders with regard to sourcing, manufacturing and delivery of products'. In today's global market and environment, the rapid pace of new product introductions and global logistics and distribution disruptions have all imposed unprecedented pressures on global SC compliance. For example, food and pharmaceutical products should be kept at the right temperature or humidity during transportation. Maintaining high quality and ensuring the safety of products requires that every segment in the logistics chain transports products in the right conditions. Due to insufficient cold transport technology capabilities, 200 million tons of food spoil before reaching market every year (Microsoft 2018).

Supporting effectively SC compliance allows for smoother operations and the uninterrupted flow of products; this involves close collaboration among various units within an organization and third-parties, including suppliers, distributors, brokers, and other SC intermediaries (e.g., 3PLs, freight forwarders, etc.). A vast number of requirements need to be monitored and adhered in order to avoid potential regulatory scrutiny and negative impacts to the organization's bottom line and reputation. A comprehensive compliance profile would consider: trade (both imports and exports), product safety and integrity, technical regulations, security (both cyber and physical), logistics and distribution, supplier integrity and social responsibility, ethical sourcing, and environmental responsibility (Deloitte 2018). The Guardian (2017) revealed that the UK's top supplier of supermarket chicken deceived customers into buying expired chicken by tampering with food safety records.

There are two fundamental questions for addressing current and emerging SC compliance issue:

- How do SC stakeholders obtain information about their compliance requirements?
- How do they coordinate and communicate compliance requirements throughout the SC to enable effective execution?

### *2.6 Trust and Stakeholder Management*

Trust is recognized as one of the most important factors in a committed and collaborative relation between SC stakeholders. For example, a manufacturer in order to be able to use properly sourced raw materials and to be transparent about its activities, is strongly dependent on its suppliers to meet factory safety standards. Trust is also essential when complying with regulatory agencies, such as customs and other governmental agencies/enforcers. Moreover, SCs need trust in order to be flexible and agile to the increased volatility they have to deal with.

Currently, SC stakeholders rely heavily on central intermediaries as brokers of trust to verify, record and coordinate transactions. For example, many of supply purchases, transactions, or transfers of funds require some form of intermediary, like a bank or a legal entity. As a result, too many middlemen and intermediaries are involved. Each of these intermediaries usually takes a cut, and can be the cause of unnecessary delays.

## 3. Blockchain Technology

Blockchain is an undeniably ingenious innovation that has created the backbone of a new type of internet and associated business models.

### *3.1 Description of Blockchain Technology*

Blockchain is a *distributed ledger* for maintaining and tracking a permanent and tamper-proof record of transactional data (Gupta 2018), where a distributed ledger is a special type of database shared, replicated, synchronized, and maintained by the participants of a decentralized network. This ledger records all the transactions and tracks assets among the involved members in a business network, such as the exchange of tangible and intangible assets or digital data. Each participant in the distributed network maintains a copy of the ledger to prevent a single point of failure and these copies are all updated and validated simultaneously. Blockchain was initially conceptualized by Satoshi Nakamoto to solve the previously unsolvable double spending problem without a middleman, which further opened up a range of new possibilities (Nakamoto 2008).

Blockchain stores data in a sequential chain of cryptographic hash-linked blocks. Each block consists of the same attributes including a block number, the hash of the current block, the hash of the previous block in the chain, transaction records, and a timestamp (see Figure 3). Data in each block is 'hashed' through *hash functions*. A hash function transforms an input of letters and numbers of *any* length into an encrypted output of *fixed* length (for example, 256 bits or 32 bytes) through a mathematical algorithm. *Hashing* is the process of applying a hash function to some data and the output of a hash function is called a *hash*. A critical characteristic of a secure hash function is that given a hash, it is mathematically and computationally infeasible to determine the input that was provided to the hash function. The order in which the transactions took place is determined jointly by the block number, the previous hash, and the current hash. The timestamp of each block determines the time at which the recorded transactions took place.

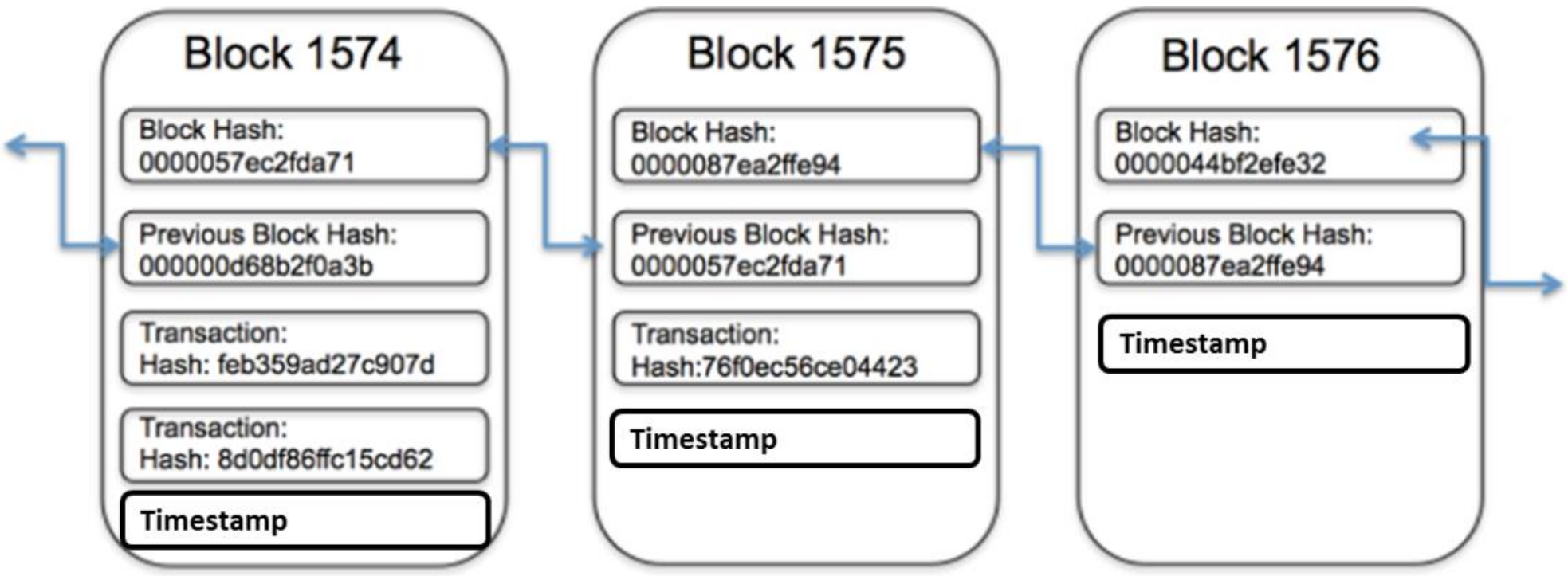

Figure 3: A representative structure of a blockchain (Gupta 2018)

The first block is created with a header and data that pertains to transactions that took place within a given time period. Afterwards, each subsequent block calculates its own hash using the previous block's hash. *Mining,* the process of adding a new block containing thousands of transactions is the most computational expensive part of the blockchain. A miner or a node is a CPU trying to guess a number (called *Nonce*) and then compute a hash based on a predescribed algorithm. If that hash is below a certain number then a new valid block is found. New blocks cannot be linked to the chain directly. The authenticity of a new block must be verified by a computational process (*validation* or *consensus)* before it can be linked to the existing chain. At this point of the blockchain process, the majority of nodes in the network must agree that the hash of the new block has been determined correctly, and thus consensus guarantees that all copies of the distributed ledger share the same state. Summing up, blockchain creates a decentralized platform where the cryptographically validated transactions and data are not under the control of any third parties. The shared ledger records transactions and data in a verifiable, transparent and permanent way. Users in a blockchain can further be confined to view only the transactions that are relevant to them.

There are several standard consensus/validation algorithms including *Proof-of-work algorithm*, *Practical byzantine fault tolerance algorithm*, *Proof-of-stake algorithm*, *Delegated proof-of-stake algorithm*. Proof of Work is especially popular as it requires high processing power to compute, but is easy to verify for other network nodes. Interested readers could see Zheng et al. (2016) and Sankar et al. (2017) for technical reviews.

### *3.2 Blockchain Platforms*

Blockchains can be public, private or permissioned, depending on who has the ability to be a user of, or run a node on, the blockchain. *Public blockchains* (e.g., Bitcoin) allow every node in the network to conduct transactions and participate in the consensus process. On the contrary, *private blockchains* (e.g., Multichain) only allow a limited number of approved nodes to participate in the consensus process. Those nodes could be a group of employees within a corporation, or a set of organizations, such as a number of banks that agree to a network.

*Permissioned blockchains* (e.g., Ripple) allow a mixture of public and private blockchains with lots of customization options. The available options include allowing anyone to join the permissioned network after a verification of its identity, and allocating selected and designated permissions to perform only certain activities on the network. Such blockchains are commonly operated by known entities such as stakeholders of a given industry.

There are many blockchain platforms depending on a variety of consensus algorithms, developing tools, and programming languages (Body 2018). Among the most important and popular blockchains are Bitcoin on crypto-currency transactions (http://bitcoin.org); Ethereum (http://ethereum.org) designed for a large variety of decentralized applications; Sidechain that can offload computations to another chain and focus on the mainchain for issues demanding the highest levels of security (Konstantopoulos 2017); Hyperledger and Hyperledger Fabric developed by IBM and Digital Asset in 2017 for business; IOTA blockchain (http://iota.org) developed to enable fee-less micro-transactions for the Internet of Things (IoT); Nebulas (https://nebulas.io) which provides a search framework for all blockchains; Skuchain (http://www.skuchain.com/) and Sweetbridge (https://sweetbridge.com/) targeted for SCM; and the Microsoft Azure ecosystem providing a cloud-based blockchain developing environment. Holotescu (2018) offers a detailed review for each of these platforms.

### *3.3 Smart Contract*

A *smart contract* is a computer program inside a blockchain containing a set of rules under which the involved parties agree to interact with each other. The agreement of a smart contract defines the conditions, rights, and obligations to which the parties of the smart contract consent. The agreement is predefined and written in digital and machine-readable form. The rights and obligations stated in the smart contract can be automatically executed by a computer or a network of computers as soon as the involved parties have reached to an agreement and satisfied the conditions of the agreement (enforcement). As the simplest form of decentralized automation, smart contracts facilitate, self-verify, and automatically enforce the negotiation and performance of an agreement. Name Bazaar (https://namebazaar.io) is using smart contracts in a peer-to-peer marketplace where buyers and sellers can directly exchange cryptographic assets on the blockchain. The advantages of smart contracts are discussed below.

- Cost savings: This is a result of the severe reduction of the elimination of vast chains of middlemen. Smart contracts establish the relationships between people, organizations and assets without the involvement of any intermediaries. The built-in code of a smart contract automatically verifies fulfilment and executes the agreed terms whenever the pre-defined rules are satisfied. Thus, smart contracts radically reduce transaction costs for reaching an agreement, formalization, and enforcement.
- Speed and accuracy: All terms and conditions of smart contracts are recorded in explicit details. Computer code is more exact and accurate than the legalese in traditional contracts, as any ambiguities or omissions in coding could result in transaction errors. Thus, smart contracts avoid the pitfalls of manually filling out stacks of paper forms. Smart contracts are automated and they self-execute and track transactions in real time, thus reducing/eliminating the time spent in processing paperwork, reconciling and correcting the errors that are often manually introduced into documents.
- Transparency and trust: The predetermined terms and conditions of smart contracts are fully visible and accessible to all relevant parties. Thus, there is no room for miscommunication or misinterpretation once the contracts are established, and nobody has to question whether a contract has been manipulated for personal benefit.

- Security and storage: All the transaction records of a smart contract are encrypted and permanently stored in a blockchain, making it really hard to modify. As all the records are linked through hashes, the whole chain would need to be alternated for changing a single transaction.

### *3.4 The Potential of Blockchain in Global SCM*

The implementation of blockchain technology can remediate the SC pain points of Figure 1. Specifically, its potential in global SCM hinges upon the following:

- The format of blockchain designs and tracking capabilities provide a full audit trail, which improves the traceability of SC networks and gives businesses increased confidence in the authenticity and quality of products. Consumers and watchdogs, either public or private, can trace every product moved through the authorized blockchain-backed platforms and validate or reject both product and involved participant. Blockchain can easily locate and correct any problems while creating an irreversible audit trail.

- The capability of blockchain in recording asset provenance, ownership transfer, legalities and safety requirements in real-time removes ambiguities and increases accountability. As such systems can provide real-time transparent data, low procedural costs, and high probability of fair judgments, disputes can be resolved in a fraction of time. Furthermore, smart contracts can automatically trigger compensations or fines if compliance with pre-set terms is violated. This makes complicated and time-consuming dispute resolution a thing of the past.

- Blockchain technology is highly secure as: (i) each participate who enters into the

blockchain network needs to have a unique identity linked to his/her account; (ii) the block encryption and how blockchain works makes it tougher for hackers to disrupt the system. Blockchain has no single point of failure and it acts as a *single source of truth*. Possession of a cargo, at each step, is tracked and monitored on the blockchain. Thus, its usage can ensure the integrity of the chain-of-custody process.

- Blockchain is becoming the backbone of SC digitization. Instead of managing bureaucracy and a lengthy paper trail, it provides an automated process of storing information in a tamper-evident digital format, with the potential to radically reduce time and costs for the transactions (Morabito 2017). The automation can further be extended to services that currently require intermediaries such as insurance, legal, brokerage, settlement services, delivery scheduling, fleet management, freight forwarding, and connectivity with business partners. Blockchain will facilitate and automate each business transaction process (e.g., automating payments and transferring ownership between parties), enabling more direct relationships among participants.

- The immutable structure of a well-designed blockchain prevents tampering and provides a reliable mechanism for SC stakeholders to prove compliance with standards. Blockchain is a more open system and data can be made readily available to auditors and other third-parties such as compliance officers. Because of its transparency and immutability, blockchain forces organizations to work within the laws and regulations to protect consumers.

- With blockchain technology, trust within a SC is easier to establish than ever before. Blockchain enables a distributed and shared environment where parties can directly conduct transactions without depending on intermediaries or any third parties to provide trust and validate the transactions. The establishment of trust is based on cryptography, distributed ledgers and consensus of blockchain. All cryptographically hashed transactions are distributed to all relevant participants. The consensus is achieved as all participants have to agree to the same version of the ledger. Thus, there is no longer a need for a central authority.

Furthermore, blockchain is expected to improve cash flow in SCs. Blockchain creates a common platform for all involved parties and facilitates improved exchange of trade information and end-to-end (E2E) transparency of the entire SC. The fast exchange of trade data and auditability of a participant's credit history can further increase speed, efficiency, and security in financing between buyers, sellers, and their banks. The real-time visibility of events along a SC implies that invoices can automatically trigger the transfer of ownership or execution of a payment, and that funds can be released faster. Blockchain can also help improve credit ratings and risk assessment procedures, ensuring security for banks, and leading to improved financing terms for both buyers and sellers (Euro Banking Association 2016). The implications of blockchain to insurance industry has been analysed in Ernst & Young (2017). By the end of September 2018, Goldman Sachs and Google Ventures have completed their participation in the latest investment round for the Veem cross-border payments start-up (Forbes 2018).

## 4. Blockchain on SCM and Global Trade: Pilot Efforts

Several pilot studies around the world and significant amount of efforts are being focused on utilizing blockchain to improve SC transparency, eliminate information delays and the uncertainty that contributes to 'the bullwhip effect', and reduce the time to resolve disputes. For example, a number of initial pilot efforts (such as that of Maersk Lines) are on blockchain-backed bills of lading to eliminate the stacks of paper documents that have long served as the foundation of global trade (American Shipper 2016, Takahashi 2016). This section further reviews a selection of leading global blockchain initiatives on global trade and SCM. We first discuss global trade initiatives, followed by efforts focused on the transportation, food, and pharmaceutical sectors. A synthesis of main foci of each pilot study is provided in Table 1.

| | Domain | Traceability | Dispute Resolution | Cargo Integrity & Security | Digitalization | Compliance | Trust & Stakeholder Management | Create Industry Standard |
|---|---|---|---|---|---|---|---|---|
| Port of Antwerp | Port | X | | X | X | | X | |
| Maersk & IBM | Container Shipping | X | X | X | X | | X | X |
| Accenture | Freight and Logistics Consortium | X | | X | X | X | | |
| BiTA | Transportation and Logistics Consortium | | | | | X | | X |
| UPS | Logistics | X | X | | X | X | X | |
| Walmart & IBM | Food | X | | | X | X | X | X |
| Carrefour | Food | X | | | X | X | X | |
| MediLedger | Pharmaceuticals | X | | | X | X | X | X |

Table 1: The main foci of pilot efforts on trade and SCM

## 4.1 *Leading Global Trade Blockchain Initiatives*

### *4.1.1 Port of Antwerp's Initiatives*

The Port of Antwerp is a vital link for global SCs (https://www.portofantwerp.com). As the second-largest seaport by total freight shipped in Europe, it is one of the fastest growing container ports of the Hamburg - Le Havre range, with cargo handled totalling 223 million tons in 2017 and projected to reach 235 million in 2018. Its shipping services cover the Americas, Africa, the Middle East and the Indian subcontinent, while it is also on its way to strengthening its position on the Far East.

The Port of Antwerp has already undertaken a first concrete step using blockchain toward a 'Smart Port', a program jointly being developed with the city of Antwerp, to become a European leader in IoT. Some of blockchain smart port applications under consideration include entirely mapping the full E2E physical flow of a container, automating the document flow, and connecting data-silos and automating joint business processes across the E2E chain. Below we discuss three initiatives that the port has undertaken.

*Secure and efficient container release.* Traditionally, a truck driver or a shipper picks up containers by a PIN code when the containers arrive in the port, ensuring that the right person picks up the right box. The container-specific PIN code is initially generated by the port terminal operator, and then is transmitted along to the forwarder and the subsequent carrier. The carrier often subcontracts the job to another haulage company and thus the PIN code has to be passed via a number of parties before it reaches the right driver. Even worse, the code is conveyed by all sorts of means including email, fax, SMS, etc. Malicious entities can simply intercept the PIN code, undermining the integrity of the SC. A legal dispute has aroused due to the usage of PIN code (Didier 2017).

To address this problem, *T-Mining*, a start-up in Antwerp has developed a blockchain solution for the port's container release operations. All necessary and required data for releasing a container are gathered in a database and all this information is restricted to the involved parties. Digital rights are created and the blockchain solution ensures that: (i) these rights can further be transferred between parties, and (ii) the sender no longer owns the right once the recipient receives it. Thus, no unauthorized entities can show up at the terminal to claim containers, except the true owner and all transactions are securely and permanently stored in the blockchain.  While traditionally the operations involve a large number of entities, the developed solution has securely digitized the operation process without any middlemen or third parties (Sluijs 2017).

*Secure and efficient document workflow.* Documents, such as certificates of origin and phytosanitary certificates, are required for the import of fruit and vegetables. The exchange of these paper-based phytosanitary certificates is usually done by post office and can involve many different parties from different countries. Given two versions of the same certificate, it can be hard to tell which certificate authentically records the origin and safety of the food. Further, transferring and processing these heaps of documents is time consuming and costly, severely impeding the SC flows.

Additionally, the Port of Antwerp and T-Mining developed a pilot project to automate and secure the flow of documents by means of smart contracts. In their pilot efforts, they focused on the transportation of apples from New Zealand to Belgium with phytosanitary certificates. The certificates were issued by the inspection authority in New Zealand, transferred by the New Zealand exporter and the Belgian importer together with the load of apples, and handed over to the Belgian authorities for inspection and approval before releasing the cargo for import in Belgium. The Port and T-Mining used blockchain to transfer these certificates to the competent authorities in Antwerp without duplicating the documents, thus guaranteeing the authenticity of the document. Smart contracts automate and secure the document flow between involved parties under pre-defined rules and all information is securely shared with the relevant parties in real time without any delay (T-Mining 2018).

*Enable chainwise collaboration.* The third pilot study by T-Mining is focusing on exchanging data among parties involved in the chain, connecting data-silos, and automating joint business processes across the E2E flow, in order to improve operational efficiency. This study is currently under development.

### *4.1.2 A Joint Venture of Maersk and IBM*

Maersk is the world's largest container shipping company. It operates in 130 countries and employs roughly 76,000 people. Maersk and IBM announced in August of 2018 that they have jointly developed the TradeLens platform that applies blockchain to global SCs. The platform empowers multiple trading participates and partners to securely share information and to collaborate by establishing a single shared view of a transaction without compromising details, privacy and confidentiality. Multiple parties can interact with each other by accessing real-time shipping data and shipping documents. ClearWay, the trade document module of TradeLens, uses smart contracts to enable collaboration in cross-organizational business processes and information exchanges in a secure and non-disputable manner. TradeLens has demonstrated its power in preventing delays caused by documentation errors, information delays, and other impediments. It has further been shown that its implementation can reduce the transit time of a shipment of packaging materials to a production line in the U.S. by 40 percent (IBM News Room 2018).

So far, more than 154 million shipping events have been recorded on the platform. These records include arrival times of vessels and container 'gate-in' and shipping documents such as bills of lading, invoices, etc. It is reported that the number of events captured on the platform is growing by one million per day. 94 organizations are actively involved or have agreed to join the platform including: more than 20 port and terminal operators globally (approximately 234 marine gateways worldwide), Pacific international carriers, customs authorities and customs brokers, cargo owners, freight forwarders, and transportation and logistics companies. TradeLens envisions using blockchain technology to (i) transform how global trade is conducted; (ii) create an industry standard for the secure digitization and transmission of SC documents; and (iii) generate tremendous savings while enhancing global SC security (IBM News Room 2018).

#### *4.1.3 Accenture's Pilot Efforts*

Accenture, a leading global management consulting and professional services firm, is also leveraging blockchain technology for secure, transparent, and efficient SC networks. In March of 2018, Accenture organized a freight and logistics consortium with members from APL, Kuehne+Nagel, AB InBev, and a European customs organization for testing a blockchain-based solution. The proposed solution was designed to eliminate the dependence on printed shipping documents and streamline the entire flow of documents related to the transportation of goods. Twelve real shipments were tested during this consortium, including beer and commodities, each of which is shipped to different destinations with its own regulatory requirements. The trials showed that the blockchain solution can cut down on inefficient data entry by as much as 80 percent. Essentially, the solution can simplify the process of updating shipping information, speed up the verification process required for cargo, and at the same time help customers avoid penalties in cases of non-compliance with customs policies (Accenture 2018a).

### 4.2 *Transportation Industry Initiatives*

The American Trucking Associations reports that nearly 71 percent of all the freight tonnage in the U.S. is shipped by trucks (American Trucking Associations 2018), with gross freight revenues from trucking reaching $676.2 billion in 2016, about 79.8 percent of the freight bill of the U.S. (American Trucking Associations 2017). Trucking is still a legacy industry, dominated by fragmentation and competition, displaying low transparency, unstandardized processes, data silos and diverse levels of technology adoption (J. Smith 2018). It is estimated that in the U.S. alone, there are roughly 1.5 million individual trucking companies employing approximately 3.5 million truck drivers. About 90 percent of these companies have six trucks or fewer (American Trucking Associations 2018). Today, in part due to the tremendous growth of ecommerce, the industry is plagued by low capacity and a lack of drivers; on the other side, truckers drive more than 29 billion miles with partial or empty truckloads.

The *Blockchain in Transport Alliance* (BiTA), funded in August 2017 by Freight Waves, is a consortium of the industry leaders in the transportation and logistics industry. The consortium encompasses manufacturers, shipping companies, and logistics technology companies focusing on blockchain education and common standards for blockchain applications in the transportation industry. Current members include UPS, Salesforce, McCleod Software, DAT, Don Hummer Trucking, Schneider, FedEx, Uber Freight, Delta Cargo and about 1,000 more applicants have applied for the membership. Use cases include visibility into truck performance history, vehicle maintenance, quality assurance, dynamic optimization, capacity monitoring, payment and pricing, fraud detection, and theft prevention. Over 85 percent of trucking-related transactions in the world are from BiTA members. By 2020, the goal of the consortium is to promote industry-wide education on the potential use cases of blockchain outside of crypto-currency, develop industry-wide standards and apply them to case studies, further encouraging early adoption within innovative start-ups and pilot programs at large organizations and corporations. At the same time, it envisions that regulatory authorities will develop auditing and compliance practices (BiTA 2018).

*United Parcel Service (UPS).* Century-old delivery giant UPS has applied for a patent that employs blockchain and distributed ledger technology to route packages throughout an international SC network involving multiple carriers (US Patent & Trademark Office 2018). The system designed by UPS can automatically determine a route based on the service offerings of a network of providers when a package is scanned into a packaging facility. As the package moves to its destination, the blockchain records all information about the shipment and verifies whether all service providers meet the obligations of their respective service offerings. The patent authors also note that smart contracts can be incorporated into the system to pay various parties within the SC network once they fulfil their obligations within a particular leg of the shipment.

### *4.3 Food Supply Chains*

Food is a huge, multi-trillion-dollar industry, while the global food SC networks encompass countless number of parties and players that are functionally and geographically diverse. This fragmented structure inhibits the free flow of information across the entire SC. When an issue occurs such as a contamination incident, low transparency may severely delay the proper investigation and the effective implementation of countermeasures. Another plague in the sector is food fraud. Organic produce, milk, coffee and tea, fruit juice, and olive oil are all on the list of commonly faked, diluted or adulterated foods (New Food Economy 2018). Consumers are demanding detailed information including where the food product was grown or produced, handled, packaged, stored, inspected, and who and which parties were involved (G. Smith 2018).

Food SC stakeholders are aiming at improving collaboration and acquiring better information on inventories, logistics, and demand forecasting, in order to: (i) optimize SC performance, and (ii) react promptly to external and disruptive events.

Blockchain has begun to be used to improve collaboration, trust, and transparency across the food industry (Mao et al. 2018, CoBANK 2018). IBM has established collaboration with multiple key players in the food industry. *Chain Business Insights* discusses how 13 large firms and start-ups in the food sector have targeted blockchain-based innovations to improve the transparency of farm-to-fork SCs (Chain Business Insights 2017). Below, we briefly discuss two relevant important large-scale initiatives worldwide.

*Walmart & IBM initiative.* Since 2016, Walmart has been working with IBM to employ and develop blockchain platforms for tracking products along farm-to-fork chains. The two companies have developed two pilot studies on food traceability and SC transparency. In the first study, they traced a package of sliced mangoes in a U.S. store back to a Mexican orchard in 2.2 seconds. The study involved 16 farms, two packing houses, three brokers, two import warehouses, and one processing facility. 23 different lot codes and tens of thousands of sliced mangoes were recorded over the 30-day period. In the past, the same exercise took them almost seven days. The mango pilot study is one of the strongest proofs of concept study within the industry to date. In the second study, Walmart and IBM also tracked several different pork products from a single supplier to local stores in China, in a joint effort with a Chinese online marketplace, JD.com (Kamath 2018). Walmart reports that a more transparent and accurate record of transactions on a blockchain could lead to benefits including safer food, enhanced flow to provide fresher products to customers, and boosted consumer trust. Just recently, the company has required all its direct suppliers of lettuce, spinach and other greens to join its food-tracking blockchain by January 31, 2019; all farmers, logistics firms and business partners of these suppliers are also mandated to join Walmart's blockchain by September 30, 2019 (Nash 2018b).

IBM and Walmart have been developing the *Food Trust Group* for improving recall, quickly identifying issues and reducing the time consumers are at risk. Dole Food Co., Driscoll's Inc., Nestle S.A., Golden State Foods, Kroger Co., McCormick and Co., McLane Co., Tyson Foods Inc, and Unilever NV are all members of the Food Trust Group aiming to set new standards for the industry. They stored data for 1M items in about 50 food categories, including Nestle canned pumpkin, Driscoll's strawberries and Tyson chicken thighs (Nash 2018a). Their effort is envisioned to encourage accountability and provide suppliers, regulators and consumers 'greater insight and transparency' into how food is handled - from farm-to-fork.

*Carrefour, the Europe's First Food Blockchain.* In March of 2018, the Europe's largest retailer launched Europe's first food blockchain with one of its iconic animal product lines: free-range Carrefour Quality Line Auvergne chicken ('Carrefour Launches Europe's First Food Blockchain'). This system is designed to guarantee complete product traceability by requiring every party along the SC --- producers, processors, and distributors--- to track their activities. Each product's label features a QR Code and shoppers can use a smartphone to scan the code on the package to obtain information of the product at each stage of production and the journey it has taken. The information includes where and how the chickens were raised, the name of the farmer, what they were fed, what treatments were used, where they were slaughtered, where the meat was processed, and when they were placed on the supermarket shelves, etc. Carrefour further rolled out blockchain technology with its quality line tomato in July, 2018. The company is aiming at extending the use of blockchain to honey, eggs, cheese, milk, oranges, tomatoes, salmon and hamburgers by the end of 2018, as it prepares a major overhaul to tackle competition from Amazon, Leclerc and others.

### *4.4 Pharmaceutical Supply Chains*

Drug counterfeiting is a well-recognized problem that affects human lives and the reputation and return on investment of the pharmaceutical industry. It is estimated that up to 30 percent of pharmaceutical products sold in emerging markets are counterfeit and about one million lives are lost due to counterfeit medication each year (World Health Organization 2018). The global counterfeit medicine business is estimated to be somewhere between $75 to $200 billion annually (GrantThornton 2018). Both the Drug Supply Chain Security Act (DSCSA) in the U.S., and the Global Traceability Standard for Healthcare (GTSH) internationally, were developed to protect consumers from counterfeit drugs. DSCSA requires that: (i) pharma distributors should be able to verify a returned product’s authenticity before they resale by 2019; and (ii) pharma companies should be able to track and trace all prescription drugs by 2023 (Enterprise Times 2018). However, so far, the full implementation of global traceability standards across pharmaceutical supply chains is still elusive.

*The MediLedger Project.* Established in 2017, it is an industry consortium with members from Block Verify, Chronicled, IBM Blockchain, FarmaTrust, iSolve, Modum, OriginTrail, Provenance, T-Mining, The LinkLab, VeChain, and Walton (Clauson et al. 2018). MediLedger intends to: (i) bring pharmaceutical manufacturers and wholesale distributors together to evaluate the potential of blockchain technology for tracking and tracing prescription medicines; and (ii) demonstrate the potential to prevent counterfeit medicines from entering the pharmaceutical SCs, while ensuring compliance with DSCSA in the U.S. MediLedger’s blockchain will store data on separate nodes or CPUs, this inhibiting data manipulation by unauthorized parities. The project will also exploit open standards and specifications operated by pharma industry stakeholders and related technology providers.

## 5. Implications of Blockchain for Customs and Governmental Agencies

With the right strategy and technology in place, customs agencies now have the opportunity to create a completely transformational cross-border trade system aiming at linking all involved stakeholders in a connected, transparent and data-rich environment. A blockchain-enabled trade system is expected to improve speed, visibility, security, and responsiveness for all participants—be they traders, customs agents or government agencies. Customs and governmental agencies finally have the chance to overcome obstacles that plagued 'Single Window' efforts before, by tapping into the IT capabilities of the private sector and its tremendous investment in blockchain.

### *5.1 Moving beyond 'Single Window'*

'Single Window', officially known as 'International Trade Data System' (ITDS), aims at streamlining the border clearance process by providing a single platform through which all shipment data and documentation are entered and managed (UNECE 2004). 'Single Window' has been promoted by several worldwide organizations associated with trade such as UNECE, World Customs Organization (WCO), and the Association of Southeast Asian Nations (ASEAN). Many countries agreed that the initiative had the potential in enhancing the implementation of standards and techniques for simplifying and facilitating information flows and information sharing between traders and governments. UNECE has been pushing countries around the globe to implement 'Single Window' since 2004 (UNECE 2004).

Over the past two decades, however, the focus of the 'Single Window' effort has been in general on providing traders with a single entry-point for submitting electronic, standardized information and document to government for customs related transactions. In practice, the systems created are often an additional layer on top of the existing customs and other governmental agency systems, and 'Single Window' actually does little to address the underlying fragmentation and complexity issues of global trade. The two main obstacles for the complete realization of 'Single Window' systems are listed below (Accenture 2018b).

- Technology: information and communication technology for truly delivering a single window had not been previously available;

- Collaboration reluctance: it is rare to find a 'Single Window' system covering all the relevant governmental authorities, agencies, and trading communities. Coordinating these various agencies and organizations (and their procedural and data requirements) into a coherent and simplified automation system has been proven challenging.

Blockchain along with IoT and cloud computing have the capability in collectively overcoming these challenges.

### *5.2 Implications for Customs*

Blockchain is poised to radically disrupt global trade. The digital ledger technology could help to reduce the huge volumes of paperwork and bureaucratic interventions necessary for legitimate trade. The implications of blockchain for customs are in the following aspects (World Customs Organization 2018):

- Data-driven and well-informed customs. Customs supported by blockchain would be able to see the necessary and accurate data associated with the cargo to be declared and keep clear track of the location and status of the cargo in real time. Complete visibility and transparency will enable data-driven decision-making processes in the daily operations of customs and other border agencies for risk analysis and targeting.

- Customs immersed in the trade process. Blockchain technology could embed customs into a common platform linking all trade-related commercial entities and further enable information sharing among all involved stakeholders. Thus, customs can cement its position as a critical node of the global trade network and could expeditiously clear cargo that has been pre-screened on its ledger without withholding them at the time of declaration. Thus, customs could optimally allocate their limited resources to cargos requiring specific scrutiny.

- Improved revenue compliance and cooperation between tax and customs. A major problem for Tax authorities is to reduce the gap between expected value-added tax (VAT) revenues and those actually collected. Due to the enhanced transparency and traceability, blockchains would make fraud and errors far easier to detect and thus reduce the gap. The same is true for the sake of customs agencies. The reliable and real-time exchange of information between customs, exporters, importers, and other related parties enhance customs' capabilities to identify fraudulent practices.

- Combating financial crimes. Criminals usually disguise their illicit proceeds by exploiting legitimate trade processes including the overvaluation or undervaluation of the goods concerned and the use of unusual shipping routes or transhipment points. Blockchain technology could be utilized to develop a network community where customs agencies and other related governmental authorities record and share information on taxpayer's trade practices and relative activities for financial transactions. This would enable relevant authorities to streamline trade finance, track events within the banking system that could be easily misused to conceal illicit financial flows, and make necessary actions in a timely, prompt and coordinated manner.

### *5.3 Relevancy to DHS and CBP in the U.S.*

Several industry partnership programs have been developed by United States Department of Homeland Security (DHS) to simplify and streamline cross-border trade, including C-TPAT (U.S. Customs and Border Protection; Cargo Security Alliance 2014) and the Container Security Initiative (CSI) (Romero 2003; Banomyong 2005). C-TPAT is a voluntary program where private sectors and officials from Customs and Border Protection (CBP) work together to improve global trade security while maintaining an efficient flow of goods. C-TPAT participants include intermodal carriers, U.S. marine port authority and terminal operators, Mexican and Canadian manufacturers, licensed U.S. customs brokers, logistics providers, exporters and importers, etc. Furthermore, DHS has to cover worldwide compliance with regional customs programs including the Authorised Economic Operator (AEO) for the European Union (similar to C-TPAT), Partners in Protection (PIP) for Canada, and international trade agreements and requirements in cargo security.

Participants of C-TPAT and CSI can improve compliance, improve adoptions of best practices, and reduce risks by adopting the blockchain technology. We summarize the benefits between before and after in Table 2.

| | Before | After |
|---|---|---|
| ***Supply Chain Collaboration*** | Fragmented environment for stakeholders involving in cross border SC and cargo shipping to collaborate on sharing cargo security information. | Shared blockchain database for frequent and timely communications among shippers, carriers, brokers and forwarders – leveraging blockchain as an information sharing/exchange and consensus platform. |
| ***Secure Chain of Custody*** | Lack of global visibility and transparency. | Chain of custody and who touches and possesses the cargo. |
| ***Carrier security protocols and communications*** | Isolated and fragmented system. | Over shared blockchain database, carrier provides in advance driver name and photo, and unique appointment or cargo release numbers. |
| ***Carrier Vetting*** | Isolated and fragmented system, vulnerable to cyber exploits and insider risks. | Carrier identity and verification based on consensus - know who is carrying your cargo (multiple carrier anchors). |
| ***Driver Vetting*** | Isolated and fragmented system, vulnerable to cyber exploits and insider risks. | Driver vetting based on consensus – multiple driver identity anchors (whitelist, DriverSafe, DriverAdvisor, employee database). |
| ***Secure Cargo Release Process*** | Vulnerability to frauds, tampering, and uniform standards on best practices. | Bring your own device and secure cargo release based on consensus. |
| ***Cyber Risks*** | Single point of failures, prone to attacks (e.g., terminal operations, IT portal, centralized database). | Improved resilience to cyber-attacks. |
| ***Insider Threats*** | Vulnerability to insider risks. | Consensus based, immutable records, much reduced insider risks. |
| ***Documentation of Compliance*** | Lack of documentation of compliance. | Documentation of compliance. Transactions are stored on the immutable blockchain database. |

Table 2: The benefits of blockchain technology for C-TPAT and CSI

A blockchain-enabled trade system would improve visibility to SC and capability to audit compliance and non-compliance of C-TPAT participants. Records of chain-of-custody ensure that those who claim to C-TPAT compliance, actually do comply. Other benefits to CBP include faster transmission of data into CBP and Automated Commercial Environment (ACE) system, as these data can be easily extracted from blockchain. DHS and CBP have started developing pilot programs using blockchain technology to improve the import/export process (Testimony of Douglas Maughan 2018).

## 6. Challenges in Adopting Blockchain in SCM

The wide adoption of blockchain technology that many expect to dramatically change the global SC market is still in its very early days. Industry experts project that on average it may take 5.9 years for the business process improvements of the distributed blockchain ledger to be widely available (CNBC 2018); see also Figure 4. Although the pilot tests initiated by leading global industry stakeholders have demonstrated the great potential of the technology, most on-going initiatives are with rather limited scope. More comprehensive efforts realising the full potential of blockchain require significant cultural, political and technical changes and the improved coordination among shippers, carriers, service providers and governmental agencies. There are several challenges that need to be overcome in order to harness its full capabilities across the SC (Morabito 2017, World Economic Forum 2018). The main ones are discussed herein.

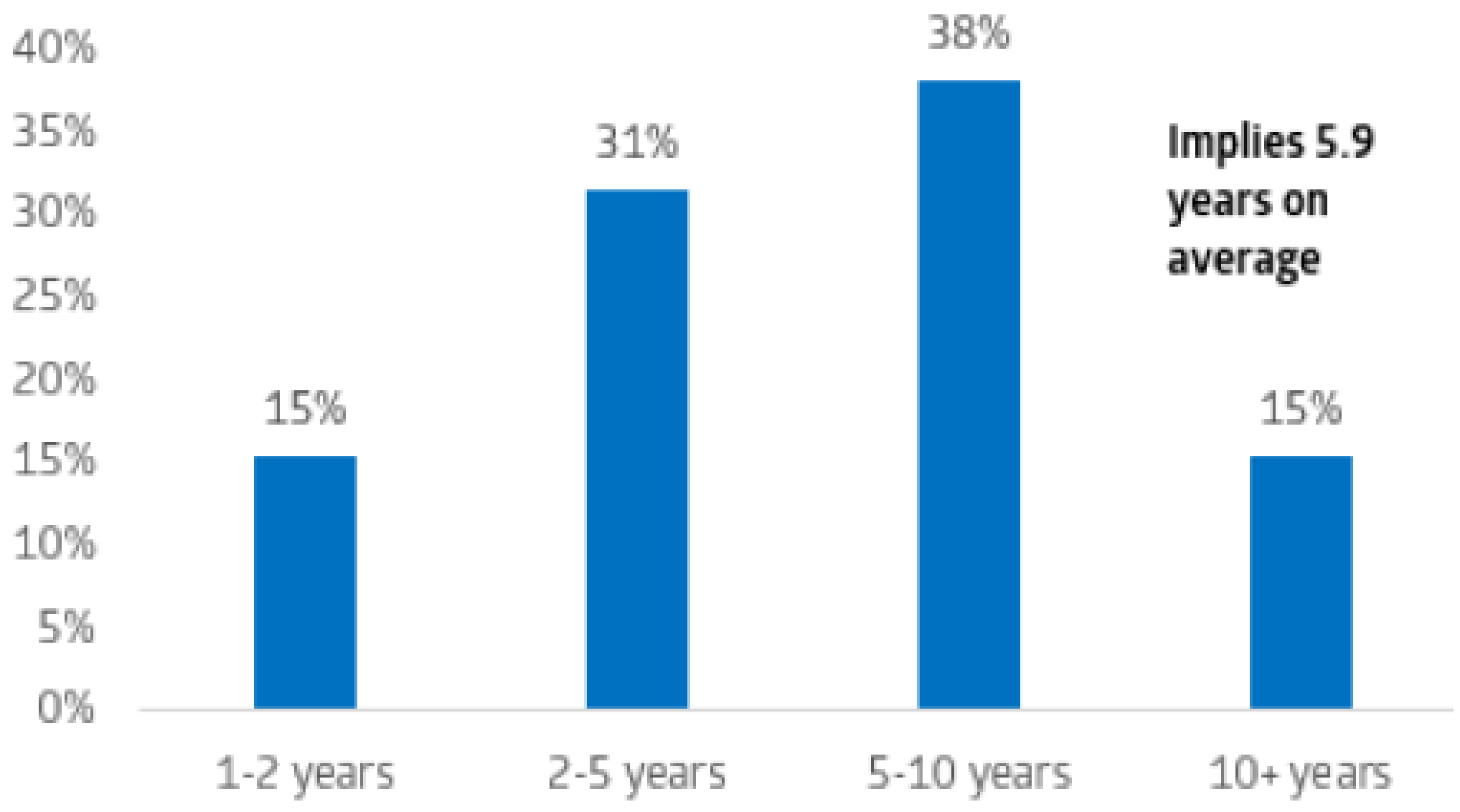

Figure 4: Expected years for widespread adoption of blockchain (CNBC 2018)

### *6.1 Adoption and Technology Challenges*

The usability of the blockchain technology is currently a crucial barrier for mainstream adoption. Lack of knowledge and broad public trust are the two main hinderances. Indeed, a sound investment on blockchain requires a certain degree of blockchain literacy. In addition, many existing interfaces for blockchain are too complex. Efforts on improving user experience, system speed and developing formalized blockchain protocols are imperative.

Scalability is required for blockchain to obtain widespread approval across industry sectors and users and to tackle global SC and trade challenges. The decentralized architecture of the blockchain network requires that every node in the network must process every transaction. Thus, blockchain applications are constrained by the time necessary to process each transaction. Today, public blockchains such as Bitcoin and Ethereum can only process three to thirty transactions per second, while 60,000 transactions per second are required for Visa (Giungato 2017). This scalability challenge due to the limited transaction capacity, will further inhibit size that the blockchain network can grow to. It is thus necessary to develop mechanisms that can limit the number of participating nodes needed to validate each transaction, while maintaining trust that each transaction is valid.

### *6.2 Interoperability*

Interoperability is the ability to easily share information, operate, and transact across various different blockchain systems. For example, users would be sending Bitcoin and receiving Ethereum naturally through blockchain interoperability, without a third party such as an exchange. Interoperability is absolutely necessary for integrating blockchain platforms with legacy systems, and with each other (Association for Financial Professionals 2017). Unfortunately, so far, interoperability has been proven elusive. In a fully interoperable environment, a user from one blockchain should be able to read, comprehend, and interact with another blockchain with little effort.

The main challenge towards achieving interoperability involves the design of a system that can relay messages between two different chains with trust. Indicatively, assume a service relays information from one chain to another; if the first chain was actually a fork of the original chain, the message being relayed to the second blockchain should essentially be invalid if the forked chain becomes orphaned (valid blocks which are not part of the main chain).

Blockchain interoperability is considered as the next major wave of innovation that may create massive value in expanding the decentralized internet. Allowing for various related blockchains to connect, it can significantly increase scalability, speed, and extensibility. There is a plethora of collaborative efforts on interoperability within the blockchain community. Two of the top projects to create a network of blockchains are *The Cosmos Network* and *The Polkadot*

*Network*. Most of these technologies propose validators as bridge between different blockchain environments (e.g., Oraclize http://www.oraclize.it/).

### *6.3 Standardization*

While it is critical to maintain the freedom for blockchain developers to be innovative, standards are required in establishing market confidence to support the roll out of blockchain technology. The more the blockchain is used, the more transactions and interactions among involved parties should be standardized. Standardization can further advance the development of blockchain by providing internationally agreed ways of working, stimulating greater interoperability, and the speedier acceptance and enhanced innovation. Many global container carriers including Maersk, APL, Kuehne+Nagel have emphasized that the future success of the shipping industry hinges upon the standardization of blockchain (Tirschwell 2018).

Many national and international organizations are working on establishing generally accepted technical rules and standards. In April 2016, Australia proposed to the International Organization for Standardization (ISO) setting up a technical committee (TC) on standardization of blockchain technology. As a result, a TC on blockchain and electronic distributed ledger technologies was established (TC 307) in September 2016, and international standardization efforts began in the areas of blockchain and electronic distributed ledger systems and applications, interoperability, and data exchange. By the end of 2017, 27 member countries have joined the TC to formalize the formation of working groups around terminology, reference architecture, taxonomy and ontology, use cases, security, privacy, and smart contracts (CISION 2017). GS1, the global business communications standards organization, has also been collaborating with IBM and Microsoft to bring barcode-like standards to blockchain-enabled systems for SC clients (GS1 2017).

### 6.4 *Binding the Physical and Digital for Complete Trust*

Binding the physical world with information stored on a blockchain is the closed loop required for complete, absolute trust. The effectiveness and efficiency of a blockchain-based SCM system relies on the assumption that information stored on blockchain correctly reflects reality. Usually, parties involved in global trade and transportation are not anonymous and are relatively stable. While an entity joining the network will be granted a private key to manage transactions, serial numbers can be easily copied and transferred, and the blockchain itself would be unaware. Truly binding the physical to the digital and creating immutable trust requires an approach that can guarantee information stored in the system can accurately reflects the status of the cargo.

On this front, recent commercial efforts and research have applied near Field Communication (NFC) chips (Chronicled, Inc. 2016), electronic tags, and IoT gadgets (Fremantle et al. 2017, Filament 2018) to securely convert physical events to digital inputs for blockchains. An authenticated data feed system was developed by Zhang et al. (2016) to ensure correctness of inputs fetched by smart contracts. Xu et al. (2018) also proposed a binding scheme leveraging a digital identify management mechanism. Their approach aims at mapping the best practice in the physical to the digital world and reducing information inconsistency.

### 6.5 *Legal and Regulatory Challenges*

Significant regulatory and legal challenges exist for the wide adoption of blockchain in global SCs. The main challenges as classified by the World Economic Forum (2018) are discussed below.

- Distributed jurisdiction and laws: As each node of a blockchain ledger is potentially located in a different part of the world, blockchain ledgers do not have a clearly identified location for each transaction. Consequently, it is not clear under which jurisdiction a blockchain will fall. Deciding which law(s) should be followed and which courts have the right to decide on what matters could be a complex and even conflicting task.
- Legal framework to ensure legal validity: For the successful deployment of smart contracts and transactions, the legal framework on contract formation and recognition should be adaptive to reflect technological developments. The blockchain should be recognized as immutable by law. Also, the legal basis for contract formation should evolve so no doubt will arise when an agreement is deemed to be valid and enforceable.
- Responsibility and accountability: By the nature of blockchain, there is no single owner of a blockchain system. Thus, knowing who should be held accountable is often unclear and attributing responsibility for blockchain technology is challenging. Legal and regulatory frameworks should clarify accountability and attribute responsibility for their actions in a sensible and timely manner.
- Data privacy: The immutability of blockchain raises the question of data privacy, especially for personal data. Cross-border blockchain platforms are examples of public networks that will handle personal data. How to balance an individual's right to privacy in an open network is a challenging task. Many blockchain networks today have little control over where data is transferred to and who has access to it.

## 7. Future Research Directions

Despite growing investment from private sectors on blockchain, few efforts have focused on the integration of the needs and requirements from governmental agencies. There is a lack of understanding of how governmental agencies across the world should be involved and how to level these initiatives led by the private sector. In addition, there is a potential tension between business operators and technology/solution providers. In the U.S., it is in the interest of CBP to promote the adoption and develop many of open architecture and standards, something that may not be always aligned with the interests of technology providers. Support from governments can ensure that a blockchain-based SC ecosystem for data sharing and information exchange encompasses requirements from both governmental agencies and the private sector.

While developing new technologies and novel business models based on blockchain, research universities and companies should work closely with governmental agencies for further facilitating global trade and ensuring compliance with trade law and regulations. This can be achieved by taking advantage of the unique characteristics of blockchain and smart contracts (e.g., consensus driven, data exchange in decentralized/distributed IT environment, immutability of history, strong protection of data integrity, cyber-attack resilience, auditability) in order to: (i) streamline and harmonize information along SC networks; (ii) improve data quality; (iii) support the timely analysis of SC risks; and (iv) develop efficient business processes between governments and global SC stakeholders.

Specifically, future research directions for joint efforts among government, industry and academia should include:

- Assessing from technology, business, policy, as well as operational aspects, where blockchain and smart contracts can be applied to the current flow of information and documents for governmental agencies. Identification of places where blockchain can be integrated for automation, data harmonization, and information exchange is a critical step for the governmental agencies to understand the business flow, benefits, and potential use cases of blockchain to secure trade.

- Developing artifacts and knowledge for governmental agencies in forms of architecture specification, analysis, and recommendations. Artifacts can include open architecture designs as well as visual representation and specifications of business flows and blockchain architecture. The analysis and recommendations should provide guidance on how government could be involved in the ecosystem. While government may not be the owner of the blockchain infrastructure, this effort may help to integrate governmental agencies as a critical participant in the SC network to facilitate trade and secure transactions. The value proposition for governmental agencies and involved parties regarding the use of blockchain in global trade should be clearly articulated and conveyed.
- Identifying both technical and non-technical challenges regarding feasibility and viability of integrating blockchain into global trade from a governmental perspective. Such issues may include confidentiality assurance, access control, interoperability, open architecture specification, and integration with standardization.
- Exploring path forward and solutions where interoperability can be achieved. Blockchains are verticalizing as they are tailored to specific business needs (e.g., finance, SCM) and people. Identification of the path forward and recommendations on this front can enhance the integration of blockchain with existing business flows and the use of international standards (e.g., WCO, W3C).

## Acknowledgement

This material is based upon work supported by the U.S. Department of Homeland Security under Grant Award Number 2015-ST-061-BSH001. This grant is awarded to the Borders, Trade, and Immigration (BTI) Institute: A DHS Center of Excellence led by the University of Houston, and includes support for the project 'Secure and Transparent Cargo Supply Chain: Enabling Chain-of-custody with Economical and Privacy Respecting Biometrics, and Blockchain Technology' awarded to University of Houston. The views and conclusions contained in this document are those of the authors and should not be interpreted as necessarily representing the official policies, either expressed or implied, of the U.S. Department of Homeland Security.